\documentclass[reprint,superscriptaddress,footinbib,aps,prd,10pt]{revtex4-1}
\usepackage{amsmath,amssymb,marvosym}
\renewcommand{\d}{\ensuremath{\mathrm{d}}}

\begin{document}

\title{{\Large {\bf The Minimal Landau Background Gauge on the Lattice}}}

\author{Attilio~Cucchieri}
\author{Tereza~Mendes}
\affiliation{Instituto de F\'\i sica de S\~ao Carlos, 
Universidade de S\~ao Paulo, \\
        Caixa Postal 369, 13560-970 S\~ao Carlos, SP, Brazil}

\begin{abstract}
We present the first numerical implementation of the minimal Landau 
background gauge for Yang-Mills theory on the lattice. Our approach is 
a simple generalization of the usual minimal Landau gauge and is formulated
for general SU(N) gauge group. We also report on preliminary tests of
the method in the four-dimensional SU(2) case, using different background 
fields. Our tests show that the convergence of the numerical minimization 
process is comparable to the case of a null background. The
uniqueness of the minimizing functional employed is briefly discussed.
\end{abstract}

\maketitle

%%%%%%%%%%%%%%%%%%%%%%%%%%%%%%%%%%%%%%%%%%%%%%%%%%%%%%%%%%%%%%%%%%%%%%%%%%%%%%%

In Ref.\ \cite{Cornwall:2009as} Cornwall pleaded with the lattice community 
for an answer to the following question:
{\em Can you find a way of doing lattice simulations in the background-field 
Feynman gauge?}
The reason for this request is that one can show \cite{Binosi:2002ft} --- to 
all orders in perturbation theory --- that there is a simple correspondence 
between the background-field method in the Feynman gauge \cite{Dashen:1980vm} 
and the so-called pinch technique \cite{Cornwall:1981zr}, which allows one to 
build gauge-invariant off-shell Green functions in the continuum.

Let us note that the numerical implementation of Landau gauge fixing (e.g.\
for the evaluation of Green functions) is well understood \cite{Giusti:2001xf}.
Recently, it has been shown that practical simulations of the linear covariant
gauge are also possible \cite{Cucchieri:2009kk} and that, with a suitable
discretization of the gluon field, it becomes feasible to treat the Feynman
gauge \cite{covariant}. Here we present the first numerical implementation of
the minimal Landau background gauge on the lattice. Our proposal is based on
Ref.\ \cite{Zwanziger:1982na}, which considers this gauge in the continuum.  

%%%%%%%%%%%%%%%%%%%%%%%%%%%%%%%%%%%%%%%%%%%%%%%%%%%%%%%%%%%%%%%%%%%%%%%%%%%%%%

\vskip 3mm

The covariant background gauge condition is introduced \cite{PS} by splitting
the (continuum) Yang-Mills field $A_{\mu}(x)$ into a quantum fluctuation
component $Q_{\mu}(x)$ and a background field $B_{\mu}(x)$, i.e.\
\begin{equation}
A_{\mu}(x) \, = \, Q_{\mu}(x) + B_{\mu}(x) \; ,
\label{eq:splitting}
\end{equation}
where $A_{\mu}(x)$ is given in terms of the generators
$T_b$ of the SU(N) gauge group by $A_{\mu}(x) = A_{\mu}^b(x) \, T_b$
[and similarly for $Q_{\mu}(x)$ and $B_{\mu}(x)$]. Note that
$B_{\mu}(x)$ is in principle arbitrary \cite{bg}.
Then, the usual covariant gauge condition
\begin{equation}
\partial_{\mu} A_{\mu}(x) \, = \, \Lambda(x) \, = \, \Lambda^b(x) \, T_b
\label{eq:covariant}
\end{equation}
becomes
\begin{equation}
\partial_{\mu} Q_{\mu}(x) \, + \, i \, [B_{\mu}(x), \,Q_{\mu}(x)] 
\, \equiv \, D_{\mu}[B] \, Q_{\mu}(x) \, = \, \Lambda(x) \; .
\label{eq:bgcovariant}
\end{equation}
Here $D_{\mu}[B]$ is the background-field covariant derivative and
$\Lambda^b(x)$ is a Gaussian-distributed real variable. Clearly, for 
a null background field $B_{\mu}(x) = 0$ one has $Q_{\mu}(x) = A_{\mu}(x)$ 
and the usual covariant gauge condition (\ref{eq:covariant}) is recovered. 
For $\Lambda(x) = 0$ the gauge condition (\ref{eq:bgcovariant}) is
the Landau background gauge condition.

Let us recall that the continuum gauge transformation of 
the Yang-Mills field, i.e.\
\begin{equation}
A^{(g)}_{\mu}(x) \, = \, g(x) \, A_{\mu}(x) \, g^{\dagger}(x) \, - \,
    i \, g(x) \, \partial_{\mu} g^{\dagger}(x) \; ,
\end{equation}
becomes
\begin{equation}
A^{(g)}_{\mu}(x) \, \approx \, A_{\mu}(x) \, + \, D_{\mu}[A] \, \gamma(x)  
\label{eq:gfinfinitesimal}
\end{equation}
if an infinitesimal gauge transformation
\begin{equation}
g(x) \, = \, \exp{[ \, - \, i \, \gamma(x) \,]} \, \approx \, \mbox{\bf$1$} 
\, - \, i \, \gamma(x)
\label{eq:ginfinit}
\end{equation}
is considered, where $\gamma(x) = \gamma^b(x) \, T_b$. [Note that, with our
notation, the generators $ T_b $ are Hermitian.
In what follows we will also employ the relations ${\rm Tr} \, T_b = 0$ and
${\rm Tr} \, \{  T_b \, T_c \} \propto \delta_{bc}$.] Then, using the
splitting in Eq.\ (\ref{eq:splitting}), there is clearly no unique way of
defining the infinitesimal gauge transformations $Q^{(g)}_{\mu}(x)$ and
$B^{(g)}_{\mu}(x)$ for the quantum fluctuation and the background fields.
Indeed, depending on which of the three terms $\,\partial_{\mu} \gamma(x)$,
$i\,\left[\,Q_{\mu}(x), \,\gamma(x)\,\right]$ and
$\, i \left[\,B_{\mu}(x),\,\gamma(x)\,\right]$ [see 
Eq.\ (\ref{eq:gfinfinitesimal})] are included in $Q^{(g)}_{\mu}(x)$ and
$B^{(g)}_{\mu}(x)$, eight different sets of gauge transformations arise
naturally. Among these, two common choices are
\begin{eqnarray}
\!\!\!\!\! Q^{(g)}_{\mu}(x) & = & Q_{\mu}(x) \,+\, D_{\mu}[B] \,\gamma(x)
\, + \, i \, \left[ \, Q_{\mu}(x), \gamma(x) \, \right]
\label{eq:gfQquantum} \\[2mm]
\!\!\!\!\! B^{(g)}_{\mu}(x) & = & B_{\mu}(x)
\label{eq:gfAquantum}
\end{eqnarray}
and
\begin{eqnarray}
\hspace{-2.5cm} Q^{(g)}_{\mu}(x) & = & Q_{\mu}(x) \,+\,  
i\,\left[\,Q_{\mu}(x), \,\gamma(x) \, \right]
\label{eq:gfQback} \\[2mm]
\hspace{-2.5cm} B^{(g)}_{\mu}(x) & = & B_{\mu}(x) \,+\, D_{\mu}[B]\,\gamma(x) \;.
\label{eq:gfAback}
\end{eqnarray}
These two transformations are referred to \cite{Pokorski} as the 
quantum transformation and the background transformation, respectively.

%%%%%%%%%%%%%%%%%%%%%%%%%%%%%%%%%%%%%%%%%%%%%%%%%%%%%%%%%%%%%%%%%%%%%%%%%%%%%%%%

\vskip 3mm

The minimal Landau gauge (in the continuum) is obtained 
\cite{Vandersickel:2012tz} by considering stationary
points of the minimizing functional
\begin{equation}
{\cal E}[A,g] \, = \, \int d^{\d}x \; {\rm Tr}
\left\{ \, A^{(g)}_{\mu}(x) \, A^{(g)}_{\mu}(x) \, \right\} \; .
\label{eq:ELandau}
\end{equation}
Indeed, the first variation with respect to the gauge transformation 
$g(x)$ gives
\begin{eqnarray}
{\cal E}[A,g] & \approx & {\cal E}[A,\mbox{\bf$1$}] \,+\, 2 \, 
\int d^{\d}x \; {\rm Tr} 
\Bigl\{ \, A_{\mu}(x) \,  D_{\mu}[A] \, \gamma(x) \, 
\Bigr\} \nonumber \\[2mm]
& = & {\cal E}[A,\mbox{\bf$1$}] \,-\, 2 \, \int d^{\d}x \; {\rm Tr} \left\{ \,
        \gamma(x) \, \partial_{\mu} A_{\mu}(x) \, \right\} \; ,
\end{eqnarray}
where we used Eq.\ (\ref{eq:gfinfinitesimal}), the relation
\begin{equation}
{\rm Tr} \Bigl\{\,A_{\mu}(x) \,\left[\,A_{\mu}(x), 
                               \,\gamma(x)\,\right]\,\Bigr\} \,=\, 0
\label{eq:Trrelation}
\end{equation}
and integration by parts. (As is usually done, we make the assumption that 
the boundary term in the integration by parts gives a null contribution.) 
Thus, a stationary point of the functional (\ref{eq:ELandau}) satisfies the 
condition 
\begin{equation}
{\rm Tr} \Bigl\{ \,T_b \, \partial_{\mu} A_{\mu}(x) \,\Bigr\} \,=\, 0 \; ,
\label{eq:Landaugauge}
\end{equation}
which is equivalent to Eq.\ (\ref{eq:covariant}) for $\Lambda(x)=0$.

Working in a similar way, one can also obtain the minimal Landau background 
gauge. Indeed, the minimization of the functional \cite{Zwanziger:1982na}
\begin{equation}
{\cal E}[Q,g] \, = \, \int d^{\d}x \; {\rm Tr} \left\{\, 
                      Q^{(g)}_{\mu}(x) \, Q^{(g)}_{\mu}(x) \, \right\} \; ,
\label{eq:EbgLandau}
\end{equation}
yields the variation
\begin{eqnarray}
{\cal E}[Q,g] & \approx & {\cal E}[Q,\mbox{\bf$1$}] \, + \, 2 \, 
     \int d^{\d}x \; {\rm Tr} \Bigl\{ \, Q_{\mu}(x) \,  D_{\mu}[B] \, 
     \gamma(x) \nonumber \\[2mm]
     & & \qquad \quad + \, i \, Q_{\mu}(x) \, \left[ \, Q_{\mu}(x), \,
     \gamma(x) \, \right] \, \Bigr\} \; ,
\end{eqnarray}
if we use the gauge transformation (\ref{eq:gfQquantum}). The above 
expression may be written as
\begin{equation}
  {\cal E}[Q,g] \, \approx \, {\cal E}[Q,\mbox{\bf$1$}] \, - \, 2 \, \int 
  d^{\d}x \; {\rm Tr} \Bigl\{\, \gamma(x) \, D_{\mu}[B] \, Q_{\mu}(x) \,\Bigr\}
\end{equation}
if we again integrate by parts, use Eq.\ (\ref{eq:Trrelation}) and
note the relation
\begin{equation}
{\rm Tr} \Bigl\{ Q_{\mu}(x)\, [B_{\mu}(x), \gamma(x)] \Bigr\} \, = \,
- {\rm Tr} \Bigl\{ \gamma(x) \, [B_{\mu}(x), Q_{\mu}(x)] \Bigr\}\;.
\end{equation} 
Thus, in this case, the stationarity condition implies the gauge-fixing 
relation
\begin{equation}
\label{eq:BGcondition}
{\rm Tr} \Bigl\{\,T_b \, D_{\mu}[B] \, Q_{\mu}(x) \,\Bigr\} \,=\, 0 \;,
\end{equation}
which is equivalent to Eq.\ (\ref{eq:bgcovariant}) for $\Lambda(x)=0$.
Clearly, for a null background, i.e.\ $B_{\mu}(x)=0$, the minimizing 
functional (\ref{eq:EbgLandau}) coincides with the usual Landau-gauge 
functional (\ref{eq:ELandau}) and the gauge condition (\ref{eq:Landaugauge}) 
is recovered.

More in general one should note that, by considering quadratic terms in 
$Q_{\mu}(x)$ and $B_{\mu}(x)$, there are only three terms that can contribute 
to the minimizing functional of the minimal Landau background gauge, i.e.\ 
$Q_{\mu}(x) \, Q_{\mu}(x)$, $Q_{\mu}(x) \, B_{\mu}(x)$ and
$B_{\mu}(x) \, B_{\mu}(x)$. However, if one wants to obtain the minimal 
Landau-gauge functional (\ref{eq:ELandau}) in the limit $B_{\mu}(x) \to 0$, 
then the minimizing functional ${\cal E}[Q,g]$ in Eq.\ (\ref{eq:EbgLandau}) 
is the only choice at our disposal. 
In this sense, the minimizing functional ${\cal E}[Q,g]$ is
{\em unique}. Moreover, of the eight natural sets of gauge transformations
for the quantum field and the background field (see discussion above), one 
can verify that only the quantum transformation 
(\ref{eq:gfQquantum})--(\ref{eq:gfAquantum}) and the set
\begin{eqnarray}
Q^{(g)}_{\mu}(x) & = & Q_{\mu}(x) \,+\, D_{\mu}[B] \, \gamma(x) \\[2mm]
B^{(g)}_{\mu}(x) & = & B_{\mu}(x) \,+\, 
                       i \,\left[\,Q_{\mu}(x), \,\gamma(x)\,\right]
\end{eqnarray}
yield the gauge condition (\ref{eq:BGcondition}). Of course, if one lifts 
the requirement of recovering the functional (\ref{eq:ELandau}) for 
$B_{\mu}(x) = 0$, then the minimal background Landau gauge can
also be implemented by considering for example the minimizing functional
$\, \int d^{\d}x \; {\rm Tr} \{\,Q^{(g)}_{\mu}(x) \, B^{(g)}_{\mu}(x)\,\} \,$
with the gauge transformation $Q^{(g)}_{\mu}(x) = Q_{\mu}(x)$ and
$ B^{(g)}_{\mu}(x) = B_{\mu}(x) \,+\, D_{\mu}[B] \, \gamma(x) \,+\,
i \,\left[\, Q_{\mu}(x), \gamma(x) \,\right]$.

%%%%%%%%%%%%%%%%%%%%%%%%%%%%%%%%%%%%%%%%%%%%%%%%%%%%%%%%%%%%%%%%%%%%%%%%%%%%%%%%

\vskip 3mm

The above results may be easily extended to the lattice formulation of 
Yang-Mills theories. To this end, we write the link variables entering 
the lattice action as \cite{Cea:1990td}
\begin{equation}
U_{\mu}(x) \, = \, W_{\mu}(x) \, V_{\mu}(x) \; .
\end{equation}
We also set
\begin{eqnarray}
\label{eq:UA}
U_{\mu}(x) & = & \exp{\left[ \, i \, a \, A_{\mu}(x) \, \right]} \\[2mm]
W_{\mu}(x) & = & \exp{\left[ \, i \, a \, Q_{\mu}(x) \, \right]} \\[2mm]
V_{\mu}(x) & = & \exp{\left[ \, i \, a \, B_{\mu}(x) \, \right]} \; ,
\label{eq:VB}
\end{eqnarray}
where $a$ is the lattice spacing. At the same time, we define 
\cite{Leinweber:1998uu}
\begin{equation}
\left. A_{\mu}(x) \,=\, \frac{U_{\mu}(x) \,-\, U^{\dagger}_{\mu}(x)}{2 i a} 
\, \right|_{traceless} \;,
\label{eq:Alatticedef}
\end{equation}
and similarly for $Q_{\mu}(x)$ and $B_{\mu}(x)$. Then, Eq.\
(\ref{eq:splitting}) is immediately recovered, modulo discretization
effects.

The lattice gauge transformation
\begin{equation}
U^{(g)}_{\mu}(x) \, = \, g(x) \, U_{\mu}(x) \, g^{\dagger}(x+a e_{\mu})
\label{eq:Ugf}
\end{equation}
can also be split among the quantum link $W_{\mu}(x)$ and the background link
$V_{\mu}(x)$. For example, the quantum transformation 
(\ref{eq:gfQquantum})--(\ref{eq:gfAquantum}) is obtained by considering
\begin{eqnarray}
\label{eq:gfQquantumlattice}
W^{(g)}_{\mu}(x) & = & g(x)\,W_{\mu}(x) \,V_{\mu}(x) \,g^{\dagger}(x+a e_{\mu})
            \, V^{\dagger}_{\mu}(x) \\[2mm]
V^{(g)}_{\mu}(x) & = & V_{\mu}(x) \,,
\label{eq:gfAquantumlattice}
\end{eqnarray}
while for the background transformation (\ref{eq:gfQback})--(\ref{eq:gfAback}) 
we have 
\begin{eqnarray}
\label{eq:gfQbacklattice}
W^{(g)}_{\mu}(x) & = & g(x) \, W_{\mu}(x) \, g^{\dagger}(x) \\[2mm]
V^{(g)}_{\mu}(x) & = & g(x) \, V_{\mu}(x) \, g^{\dagger}(x+a e_{\mu}) \;.
\label{eq:gfAbacklattice}
\end{eqnarray}
Clearly, in both cases the link variable $U_{\mu}(x)$ transforms as in 
Eq.\ (\ref{eq:Ugf}). Moreover, using Eqs.\ (\ref{eq:UA})--(\ref{eq:VB}) and 
the lattice definitions of the fields $A_{\mu}(x)$, $Q_{\mu}(x)$ and 
$B_{\mu}(x)$ in terms of the link variables $U_{\mu}(x)$, $W_{\mu}(x)$
and $V_{\mu}(x)$, one recovers Eqs.\ (\ref{eq:gfQquantum})--(\ref{eq:gfAback}) 
when an infinitesimal gauge transformation (\ref{eq:ginfinit}) is considered. 
For example, Eq.\ (\ref{eq:gfQquantumlattice}) gives
\begin{widetext}
\begin{eqnarray}
W^{(g)}_{\mu}(x) & \approx & \left[ \mbox{\bf$1$} \, - \, i \, \gamma(x) \right]
            \, \left[ \mbox{\bf$1$} \, + \, i \, a Q_{\mu}(x) \right]
            \, \left[ \mbox{\bf$1$} \, + \, i \, a B_{\mu}(x) \right]
            \, \left[ \mbox{\bf$1$} \, + \, i \, \gamma(x+a e_{\mu}) \right]
            \, \left[ \mbox{\bf$1$} \, - \, i \, a B_{\mu}(x) \right] \\[2mm]
                 & \approx & \mbox{\bf$1$} \,+\, i a \Bigl\{ \, 
                      \partial_{\mu} \gamma(x) \,+\, Q_{\mu}(x)
                      \, + \, i \, [ Q_{\mu}(x), \gamma(x) ]
                      \, + \, i \, [ B_{\mu}(x), \gamma(x) ] \, \Bigr\}
                 \, = \, \mbox{\bf$1$} \,+\, i a Q^{(g)}_{\mu}(x) \; ,
\end{eqnarray}
\end{widetext}
in agreement with Eq.\ (\ref{eq:gfQquantum}).

%%%%%%%%%%%%%%%%%%%%%%%%%%%%%%%%%%%%%%%%%%%%%%%%%%%%%%%%%%%%%%%%%%%%%%%%%%%%%%%%

\vskip 3mm

One can also define a minimizing functional for the Landau background gauge on 
the lattice. Indeed, in the limit of small lattice spacing $a$, the functional
\begin{equation}
{\cal E}[W,g] \, = \, - \, \sum_{x, \mu} \, \Re \, {\rm Tr} W^{(g)}_{\mu}(x)
\label{eq:EbgLandaulattice}
\end{equation}
is equivalent to
\begin{equation}
 {\cal E}[W,g] \, \approx \, a^2 \, \sum_{x, \mu} \, {\rm Tr}
                 \Bigl\{ \, Q^{(g)}_{\mu}(x) \, Q^{(g)}_{\mu}(x) \, \Bigr\} \; ,
\end{equation}
modulo constant terms. (Here we use $\Re$ to indicate the real part.)
At the same time, for $V_{\mu}(x) = \mbox{\bf$1$}$ and
$W^{(g)}_{\mu}(x) = U^{(g)}_{\mu}(x)$ we recover the usual minimizing functional
for the Landau-gauge condition \cite{Giusti:2001xf}
\begin{equation}
{\cal E}[U,g] \, = \, - \, \sum_{x, \mu}
              \, \Re \, {\rm Tr} \Bigl\{ \, g(x) \, U_{\mu}(x) \,
                            g(x+a e_{\mu}) \, \Bigr\} \; .
\label{eq:ELandaulattice}
\end{equation}
Also, if $W^{(g)}_{\mu}(x)$ transforms as in Eq.\ (\ref{eq:gfQquantumlattice})
and we consider an infinitesimal gauge transformation (\ref{eq:ginfinit}) we
find
\begin{eqnarray}
{\cal E}[W,g] & \approx &
{\cal E}[W,1] \,-\, i \,\sum_{x, \mu} \, \Im \,{\rm Tr} \Bigl\{\,\gamma(x) \,
              \bigl[ \, U_{\mu}(x) \, V^{\dagger}_{\mu}(x)  \nonumber \\[2mm]
              & & \qquad \quad \, - \,
              V^{\dagger}_{\mu}(x-a e_{\mu}) \, U_{\mu}(x-a e_{\mu})
                                                   \, \bigr] \, \Bigr\} \; , 
\end{eqnarray}
where $\Im$ indicates the imaginary part. As a consequence, a stationary 
point of the minimizing functional (\ref{eq:EbgLandaulattice}) implies the 
gauge condition
\begin{widetext}
\begin{equation}
\, = \,{\rm Tr} \Bigl\{\, T_b \, \sum_{\mu} \,
               \left[ \, W_{\mu}(x) \, - \, W^{\dagger}_{\mu}(x) \, - \,
               V^{\dagger}_{\mu}(x-a e_{\mu}) \, U_{\mu}(x-a e_{\mu}) \, + \,
               U^{\dagger}_{\mu}(x-a e_{\mu}) \, V_{\mu}(x-a e_{\mu})
                                 \, \right] \, \Bigr\} \; ,
\label{eq:stationaritylattice}
\end{equation}
where we used the Hermiticity of the generators $T_b$. Finally, by adding 
and subtracting
$ {\rm Tr} \{ \, T_b \, \sum_{\mu} \, [ \,W_{\mu}(x-a e_{\mu}) \, - \,
W^{\dagger}_{\mu}(x-a e_{\mu}) \, ] \, \}$ we find that the null quantity
in the above equation can be written conveniently as the sum of two terms.
The first one is taken as $ {\rm Tr} \{ \, T_b \, \sum_{\mu} \, [ \,
             W_{\mu}(x) \, - \, W^{\dagger}_{\mu}(x) \, - \,
             W_{\mu}(x-a e_{\mu}) \, + \, W^{\dagger}_{\mu}(x-a e_{\mu}) 
            \, ] \, \} $ and is equal (at leading order in the
lattice spacing $a$) to
\begin{equation}
2 \, i \, a \, {\rm Tr} \Bigl\{ \, T_b \, \sum_{\mu} \,
                       \left[ \, Q_{\mu}(x) \, - \, Q_{\mu}(x-a e_{\mu}) 
                          \, \right] \, \Bigr\} \, \approx \,
2 \, i \, a^2 \, {\rm Tr} \Bigl\{ \, T_b \, \sum_{\mu} \,
                          \partial_{\mu} \, Q_{\mu}(x) \,
                          \Bigr\} \; .
\label{eq:divQ}
\end{equation}
The second term is then given by
\vskip 2mm
\begin{eqnarray}
  && {\rm Tr} \Bigl\{ \, T_b \, \sum_{\mu} \, \left[ \,
    U_{\mu}(x-a e_{\mu}) \, V^{\dagger}_{\mu}(x-a e_{\mu}) \, + \,
    U^{\dagger}_{\mu}(x-a e_{\mu}) \,V_{\mu}(x-a e_{\mu}) \nonumber \right. \\
  && \left. \qquad \qquad \qquad \qquad \, - \,
    V^{\dagger}_{\mu}(x-a e_{\mu}) \, U_{\mu}(x-a e_{\mu}) \, - \,
    V_{\mu}(x-a e_{\mu})\,U^{\dagger}_{\mu}(x-a e_{\mu}) \,\right]\,\Bigr\} \;.
\label{eq:secondquantiy}
\end{eqnarray}
\end{widetext}
Note that for a null background field, i.e.\ $B_{\mu}(x)=0$ and 
$V_{\mu}(x) = \mbox{\bf$1$}$, the quantity above is identically zero. 
In this case, we have $Q_{\mu}(x) = A_{\mu}(x)$ [i.e.\
$W_{\mu}(x) = U_{\mu}(x)$] and the gauge condition 
(\ref{eq:stationaritylattice}) becomes [see also Eq.\ (\ref{eq:divQ})] the 
usual lattice Landau-gauge condition
$ {\rm Tr} \{\, T_b \,\sum_{\mu} \,[\,A_{\mu}(x) \,-\, A_{\mu}(x-a e_{\mu})\,] 
\, \} \,=\, 0 $.
In the $B_{\mu}(x)\neq0$ case and in the limit of small lattice 
spacing $a$, one can check that the quantity (\ref{eq:secondquantiy}) is, at 
leading order, equal to the expression
\begin{equation}
 - 2 \, a^2 \, {\rm Tr} \{\,T_b \, \sum_{\mu} \,[\, B_{\mu}(x), 
                                      \,Q_{\mu}(x)\,] \,\} \;.
\end{equation}
Thus, the stationarity condition (\ref{eq:stationaritylattice}) implies 
(again at leading order in $a$)
\begin{equation}
{\rm Tr} \Bigl\{ \, T_b \, \sum_{\mu} \,
         \partial_{\mu} \, Q_{\mu}(x) \, + \,
         i \, [ \, B_{\mu}(x), Q_{\mu}(x)\, ]
         \, \Bigr\} \, = \, 0 \; ,
\end{equation}
in agreement with Eq.\ (\ref{eq:BGcondition}).

%%%%%%%%%%%%%%%%%%%%%%%%%%%%%%%%%%%%%%%%%%%%%%%%%%%%%%%%%%%%%%%%%%%%%%%%%%%%%%

\vskip 3mm

As discussed above, given a fixed lattice configuration $\{ U_{\mu}(x) \}$, 
the usual minimal Landau gauge may be imposed by numerically minimizing the 
functional (\ref{eq:ELandaulattice}).  
In particular, by considering local updates for the gauge-fixing 
transformation $\{g(x)\}$ it is easy to verify that, for a given site $y$, 
the contribution of $g(y)$ to the minimizing functional may be written as 
\cite{Cucchieri:1995pn}
\begin{equation}
{\cal E}[U,g] \, = \, \mbox{constant} \, + \, \Re \, {\rm Tr}
\Bigl\{ \, g(y) \, h(x) \, \Bigr\} \label{eq:Elocal}
\end{equation}
with $ h(x) = \sum_{\mu} [ U_{\mu}(x) \, + \, U^{\dagger}_{\mu}(x-a e_{\mu})
\, + \, U^{\dagger}_{\mu}(x) \, + \, U_{\mu}(x-a e_{\mu}) ] $.
Then, different gauge-fixing algorithms correspond to different choices for 
the iterative updates of the gauge transformation $g(y)$ in Eq.\
(\ref{eq:Elocal}).

In the case of the minimal Landau background gauge, one can consider the
minimizing functional $ {\cal E}[W,g] $, defined in Eqs.\ 
(\ref{eq:EbgLandaulattice}) and e.g.\ (\ref{eq:gfQquantumlattice}), where 
$\{ W_{\mu}(x) \}$ and $\{ V_{\mu}(x) \}$ are given (i.e.\ fixed)
quantum and background configurations respectively. It is important to 
stress that also in this case the contribution of $g(y)$ to the minimizing 
functional ${\cal E}[W,g]$ may be written as in
Eq.\ (\ref{eq:Elocal}). In this case, the quantity $h(x)$ is equal to
\begin{eqnarray}
\!\!\!\!\!\!\!\! h(x) & = & \sum_{\mu} \, \Bigl[ \, W_{\mu}(x) \, + \,
      U^{\dagger}_{\mu}(x-a e_{\mu}) \, V_{\mu}(x-a e_{\mu}) \nonumber \\[2mm]
     &   & \; + \,
      W^{\dagger}_{\mu}(x) \, + \,
      V^{\dagger}_{\mu}(x-a e_{\mu}) \, U_{\mu}(x-a e_{\mu}) \, \Bigr] \; .
\end{eqnarray}
Thus, all formulae used for the minimal Landau background gauge are natural 
generalizations of the formulae used for the usual minimal Landau gauge. 
This implies that, at least for sufficiently smooth background
configurations $\{ V_{\mu}(x) \}$, we should expect similar convergence 
of the gauge-fixing algorithms for these two gauge-fixing conditions.

In order to verify this, we have carried out some tests in the SU(2) case,
considering lattice volumes $V = 8^4$ and $V = 16^4$ with a lattice coupling
$\beta = 2.2$, corresponding to a lattice spacing $a$ of about 0.210 fermi.
This means that the thermalized configurations $\{ U_{\mu}(x) \}$ are
reasonably ``rough'' and provide a good test for the gauge-fixing
algorithm employed.
For the background-field $\{ V_{\mu}(x) \}$ we have considered three types
of configurations with three setups each, namely [here, $\sigma_j$
are the three Pauli matrices, with $\sigma_3$ being the diagonal one]:
\begin{itemize}
\item[$a)$] {\bf random center configuration} (RCC)
$V_{\mu}(x) = \pm \mbox{\bf$1$}$, which can be interpreted as a 
random configuration of {\em thin} vortices \cite{Greensite}, with, on average,
10\%, 30\% or 50\% of the links equal to $- \mbox{\bf$1$}$;
\item[$b)$] {\bf random Abelian configuration} (RAC)
$V_{\mu}(x) = \exp{[i \, \theta(x) \, \sigma_3]}$,
which may be interpreted as a random configuration of Abelian monopoles
\cite{Chernodub:1997ay}, with the angle $\theta(x)$ uniformly distributed in
the interval $[0, 2 \pi f]$ and $f$ equal to 0.1, 0.3 or 0.5;
\item[$c)$] {\bf super-instanton configuration} (SIC)
\cite{Patrascioiu:1994ei} given by
$ V_2(x) = \exp{[i \,c \,\min(x_1, N-x_1) \,\sum_j \,\sigma_j / \sqrt{3 N^2}]}$
and $ V_{\mu}(x) = 0 $ otherwise, with $c = 0.01$, 0.05 or
0.1, where $N$ is the number of lattice sites per direction.
\end{itemize}

For the two lattice volumes above, we consider ten gauge-field
configurations and, in each case, we fix the minimal background 
Landau gauge, using the stochastic-overrelaxation algorithm
\cite{Cucchieri:1995pn}, for the nine choices of background fields
described above.
The number of minimizing sweeps necessary to achieve the prescribed
accuracy was then compared to that used in the case of a null
background (i.e.\ Landau gauge). Here we stop the gauge-fixing
algorithm when the average magnitude squared of the quantity on the
r.h.s.\ of Eq.\ (\ref{eq:stationaritylattice}) is smaller than $10^{-14}$.
Note that we tuned the stochastic-overrelaxation algorithm in the case of
a null background, setting the parameter $p$ of the algorithm
(see \cite{Cucchieri:1995pn})
equal to 0.83 for $V = 8^4$ and to 0.91 for
$V = 16^4$. The same setup was then used for non-zero backgrounds.
Results of these tests are shown in Table \ref{tab}. One sees that
the convergence of the gauge-fixing algorithm for a non-zero background
is indeed similar to the case of the usual minimal Landau gauge.
Of course, by tuning the parameter $p$ also in the general case, one can
improve the results. In fact, e.g.\ for $V = 8^4$ and background RCC 30\%, we
find that with $p = 0.92$ the number of sweeps decreases considerably,
being between 418 and 653, with an average value of about 460. Similarly,
for $V = 16^4$ and the SIC background with $c=0.01$, we obtain for
$p = 0.96$ that the number of sweeps is between 794 and 1934, with an
average value of about 1001.

\setlength{\tabcolsep}{4pt}
\begin{table}[t]
\begin{tabular}{|c|c|c|c|c|c|c|} \hline
\multicolumn{1}{|c|}{$B_{\mu}(x)$} & \multicolumn{3}{c|}{$8^4$} &
                                 \multicolumn{3}{c|}{$16^4$} \\ \hline
               & aver.\ & min.\ & max.\ & aver.\ & min.\ & max.\ \\ \hline
null background  & 217 & 190 & 290 & 508 & 396 & 773 \\ \hline
RCC     10\%     & 348 & 190 & 685 & 976 & 503 &1729 \\ \hline
RCC     30\%     & 624 & 342 &1391 & 1344 & 818 & 1979 \\ \hline
RCC     50\%     & 647 & 444 &1032 & 1711 &1002 & 2714 \\ \hline
RAC     $f=0.1$  & 224 & 191 & 323 & 677 & 417 &1226 \\ \hline
RAC     $f=0.3$  & 326 & 190 &1112 & 582 & 436 & 967 \\ \hline
RAC     $f=0.5$  & 401 & 279 & 595 & 813 & 494 & 1495 \\ \hline
SIC     $c=0.01$ & 637 & 372 & 855 & 1852 &1238 & 3503 \\ \hline
SIC     $c=0.05$ & 188 & 172 & 256 & 520 & 344 & 808 \\ \hline
SIC     $c=0.1\,$& 177 & 170 & 203 & 365 & 343 & 430 \\ \hline
\end{tabular}
\caption{Average, minumum and maximum number of sweeps necessary to achieve
the prescribed accuracy for the two lattice volumes and for the nine different
background fields considered in our tests (see description in the text). For a
comparison, we also include the case of a null background.}
\label{tab}
\end{table}

%%%%%%%%%%%%%%%%%%%%%%%%%%%%%%%%%%%%%%%%%%%%%%%%%%%%%%%%%%%%%%%%%%%%%%%%%%%%%%%

\vskip 3mm

The above results indicate that numerical simulations in the 
minimal Landau background gauge are indeed feasible. One should also stress 
that the extension of the method presented here to the case of the minimal 
covariant background gauge is, in principle, straightforward 
\cite{Cucchieri:2009kk}. This extension, as well as the numerical evaluation
of Green functions in minimal Landau background gauge, is postponed to
future studies.

%%%%%%%%%%%%%%%%%%%%%%%%%%%%%%%%%%%%%%%%%%%%%%%%%%%%%%%%%%%%%%%%%%%%%%%%%%%%%%%

\vskip 3mm

\noindent
{\bf Acknowledgments:} the authors thank Daniele Binosi, Mike Cornwall and
Andrea Quadri for useful discussions. We also thank the Brazilian funding
agencies CNPq and Fapesp for partial support.

%%%%%%%%%%%%%%%%%%%%%%%%%%%%%%%%%%%%%%%%%%%%%%%%%%%%%%%%%%%%%%%%%%%%%%%%%%%%%%%


\begin{thebibliography}{99}

\bibitem{Cornwall:2009as} J.~M.~Cornwall,
   %``Open issues in confinement, for the lattice and for center vortices,''
   PoS {\bf QCD-TNT09}, 007 (2009).
   % [arXiv:0911.0024 [hep-ph]].

\bibitem{Binosi:2002ft} D.~Binosi and J.~Papavassiliou,
   %``The Pinch technique to all orders,''
   Phys.\ Rev.\ {\bf D66}, 111901 (2002).
   % [hep-ph/0208189].

\bibitem{Dashen:1980vm} R.~F.~Dashen and D.~J.~Gross,
   %``The Relationship Between Lattice and Continuum Definitions of 
   %the Gauge Theory Coupling,''
   Phys.\ Rev.\ {\bf D23}, 2340 (1981), also in
   {\em Lattice gauge theories and Monte Carlo simulations},
   C.~Rebbi (World Scientific Pub.\ Co., 1983).

\bibitem{Cornwall:1981zr} J.~M.~Cornwall,
   %``Dynamical Mass Generation in Continuum QCD,''
   Phys.\ Rev.\ {\bf D26}, 1453 (1982).

\bibitem{Giusti:2001xf} See for example Section 3
   in L.~Giusti {\it et al.}, % M.~L.~Paciello, C.~Parrinello, S.~Petrarca and B.~Taglienti,
   %``Problems on lattice gauge fixing,''
   Int.\ J.\ Mod.\ Phys.\ {\bf A16}, 3487 (2001).
   %[hep-lat/0104012].

\bibitem{Cucchieri:2009kk} A.~Cucchieri, T.~Mendes and E.~M.~S.~Santos,
   %``Covariant gauge on the lattice: A New implementation,''
   Phys.\ Rev.\ Lett.\ {\bf 103}, 141602 (2009).
   % [arXiv:0907.4138 [hep-lat]].

\bibitem{covariant}
%\bibitem{Cucchieri:2011aa}
   A.~Cucchieri, T.~Mendes, G.~M.~Nakamura and E.~M.~S.\ Santos,
   %``Feynman gauge on the lattice: New results and perspectives,''
   AIP Conf.\ Proc.\ {\bf 1354}, 45 (2011);
   %[arXiv:1101.5080 [hep-lat]].
%\bibitem{Cucchieri:2011pp} A.~Cucchieri, T.~Mendes, G.~M.~Nakamura and 
%   E.~M.~S.\ Santos,
   %``Gluon Propagators in Linear Covariant Gauge,''
   {\it ibid.} PoS {\bf FACESQCD}, 026 (2010).
   %[arXiv:1102.5233 [hep-lat]].

\bibitem{Zwanziger:1982na} D.~Zwanziger,
   %``Nonperturbative Modification Of The Faddeev-popov Formula And 
   % Banishment Of The Naive Vacuum,''
   Nucl.\ Phys.\ {\bf B209}, 336 (1982).

\bibitem{PS} See for example Section 16.6
  in {\em An Introduction To Quantum Field Theory},
  M.~E.~Peskin, D.~V.~Schroeder (Addison-Wesley Pub.\ Co., 1995).

\bibitem{bg}
%\cite{Abbott:1983zw}
%\bibitem{Abbott:1983zw} 
  L.~F.~Abbott, M.~T.~Grisaru and R.~K.~Schaefer,
  %``The Background Field Method and the S Matrix,''
  Nucl.\ Phys.\ {\bf B229}, 372 (1983);
  %%CITATION = NUPHA,B229,372;%%
%\cite{Luscher:1995vs}
%\bibitem{Luscher:1995vs} 
  M.~Luscher and P.~Weisz,
  %``Background field technique and renormalization in lattice gauge theory,''
  Nucl.\ Phys.\ {\bf B452}, 213 (1995).
  % [hep-lat/9504006].
  %%CITATION = HEP-LAT/9504006;%%

\bibitem{Pokorski} See e.g.\ Section 8.2 in {\em Gauge Field
  Theories}, S.~Pokorski (Cambridge University Press, second edition, 2000).

\bibitem{Vandersickel:2012tz} See for example Section 2.2.1
   in N.~Vandersickel and D.~Zwanziger,
   %``The Gribov problem and QCD dynamics,''
   arXiv:1202.1491 [hep-th].

\bibitem{Cea:1990td} This is a natural definition of a background field 
   configuration on the lattice [see for example P.~Cea and L.~Cosmai,
   %``Constant background fields and unstable modes on the lattice,''
   Phys.\ Lett.\ {\bf B264}, 415 (1991)]
   but, of course, other discretizations are possible.
 
\bibitem{Leinweber:1998uu} In order to reduce discretization effects --- see 
   for example D.~B.~Leinweber, J.~I.~Skullerud, A.~G.~Williams and 
   C.~Parrinello [UKQCD Collaboration],
   %``Asymptotic scaling and infrared behavior of the gluon propagator,''
   Phys.\ Rev.\ {\bf D60}, 094507 (1999)
   [Erratum-ibid.\ {\bf D61}, 079901 (2000)] ---
   one should define the r.h.s.\ of Eq.\ (\ref{eq:Alatticedef}) equal to
   $A_{\mu}(x+a e_{\mu}/2)$, where $e_{\mu}$ is a unit vector in the 
   positive $\mu$ direction, instead of $A_{\mu}(x)$. However, since the 
   leading order results coincide in the two cases, here we prefer to 
   simplify the notation and use the definition (\ref{eq:Alatticedef}).

\bibitem{Cucchieri:1995pn} A.~Cucchieri and T.~Mendes,
   %``Critical slowing down in SU(2) Landau gauge fixing algorithms,''
   Nucl.\ Phys.\ {\bf B471}, 263 (1996).
   % [hep-lat/9511020].

\bibitem{Greensite} See e.g.\ J.~Greensite,
   %``An Introduction To The Confinement Problem''
   Lect.\ Notes Phys.\ {\bf 821}, 1 (2011).

\bibitem{Chernodub:1997ay} See for example M.~N.~Chernodub and M.~I.~Polikarpov,
   %``Abelian projections and monopoles,''
   In {\em Cambridge 1997, Confinement, duality, and nonperturbative aspects 
   of QCD}, 387.

\bibitem{Patrascioiu:1994ei} A.~Patrascioiu and E.~Seiler,
   %``Superinstantons in gauge theories and troubles with perturbation theory,''
   Phys.\ Rev.\ Lett.\ {\bf 74}, 1924 (1995).
   %[hep-lat/9402003].

\end{thebibliography}
\end{document}